\documentclass[12pt]{article}
\usepackage{amsmath}
\usepackage{graphicx}

\def\papertitlepage{\baselineskip 3.5ex\thispagestyle{empty}}
\def\preprinumber#1#2{\hfill\begin{minipage}{4.2cm} #1
        \par\noindent #2 \end{minipage}}
\allowdisplaybreaks[3]

\begin{document}

\papertitlepage
\setcounter{page}{0}
\preprinumber{KEK-TH-1626}{SNUTP13-002}
\baselineskip 0.8cm
\vspace*{2.0cm}

\begin{center}
{\Large\bf Soft gravitational effects\\
in\\
Kadanoff-Baym approach}
\end{center}

\begin{center}
Hiroyuki K{\sc itamoto}$^{1)}$\footnote{E-mail address: kitamoto@post.kek.jp} and
Yoshihisa K{\sc itazawa}$^{2),3)}$\footnote{E-mail address: kitazawa@post.kek.jp}\\
\vspace{5mm}
$^{1)}${\it Department of Physics and Astronomy\\
Seoul National University, 
Seoul 151-747, Korea}\\
$^{2)}${\it KEK Theory Center, 
Tsukuba, Ibaraki 305-0801, Japan}\\
$^{3)}${\it Department of Particle and Nuclear Physics\\
The Graduate University for Advanced Studies (Sokendai)\\ 
Tsukuba, Ibaraki 305-0801, Japan}
\end{center}

\vskip 5ex
\baselineskip = 2.5 ex

\begin{center}{\bf Abstract}\end{center}
In de Sitter space, the gravitational fluctuation at the super-horizon scale may make physical quantities time dependent by breaking the de Sitter symmetry. 
We adopt the Kadanoff-Baym approach to evaluate soft gravitational effects in a matter system at the sub-horizon scale. 
This investigation proves that only the local terms contribute to the de Sitter symmetry breaking at the one-loop level. 
The IR singularities in the non-local terms cancel after summing over degenerate states between real and virtual processes.
The corresponding IR cut-off is given by the energy resolution like QED. 
Since the energy resolution is physical and independent of cosmic evolution, the non-local contributions do not induce the de Sitter symmetry breaking.  
We can confirm that soft gravitational effects preserve the effective Lorentz invariance.  

\vspace*{\fill}
\noindent
October 2013

\newpage
\section{Introduction}\label{Introduction}
\setcounter{equation}{0}

In quantum field theories in de Sitter (dS) space, 
it is well known that the propagator for a massless and minimally coupled scalar field does not respect the full dS symmetry \cite{Vilenkin1982,Linde1982,Starobinsky1982}. 
Such a symmetry breaking is caused by the fact that the propagator is sensitive to the increasing degrees of freedom at the super-horizon scale 
and so we need to introduce an infra-red (IR) cut-off. 
The IR cut-off is an initial size of universe when the exponential expansion starts 
and gives the propagator a logarithmic dependence of the scale factor of universe: $\log a(\tau)$.  
In some field theoretic models in dS space, physical quantities may acquire growing time dependences through the propagator. 

On the dS background, some modes of gravity behave just like massless and minimally coupled scalar fields. 
Unlike a scalar field, the gravitational field is massless without fine-tuning the action. 
Thus it is an attractive candidate to make physical quantities time dependent \cite{Tsamis1992}. 

Since we need to introduce all possible counter terms to renormalize ultra-violet (UV) divergences in quantum gravity, 
there are infinite choices of finite UV contributions.  
However there is no ambiguity in investigating time dependences of physical quantities. 
Concerning the internal loop contributions, we can separate IR contributions and UV contributions in comparison to the Hubble scale. 
The degrees of freedom at the sub-horizon scale are constant with cosmic evolution and so the UV contributions respect the dS symmetry.    

In the previous studies \cite{KitamotoSD}, 
we investigated soft gravitational effects on matter systems at the sub-horizon scale which are directly observable. 
As specific examples, we adopted massless and conformally coupled scalar and massless Dirac fields. 
Since dS space is conformally flat, the matter actions possess the Lorentz invariance at the classical level after the conformal transformation.
In a generic non-conformally coupled case, the conformal invariance holds at the sub-horizon scale 
and the Lorentz invariance appears as an effective symmetry. 
We found that the effective Lorentz invariance is preserved even if soft gravitational effects are considered. 

Here we recall that in massless field theories, IR singularities occur even in flat space when virtual particles approach on-shell. 
Such singularities originate in the fact that the integration over the infinite past is divergent when the frequency of the integrand vanishes. 
In QED and QCD, these IR singularities are known to cancel after summing over degenerate states between real and virtual processes. 
The corresponding IR cut-off is given by the energy resolution \cite{Kinoshita1962,Lee1964}. 
We found that the cancellation takes place in $\varphi^3$, $\varphi^4$ theories in dS space \cite{Kitamoto2010}. 
The energy resolution is physical and independent of cosmic evolution. 
Therefore these non-local contributions do not induce the dS symmetry breaking. 
It would be appropriate to postulate that the cancellation takes place in any unitary model as the total spectral weight is preserved. 

From the above discussion, we need to distinguish the dS symmetry breaking which is local from the non-local contribution. 
In the previous studies, we adopted the effective equation of motion \cite{Hu1997} as a tool to evaluate time dependent quantum corrections.  
There we have shown that the non-local contribution does not lead to the dS symmetry breaking in the off-shell case. 

In order to investigate the on-shell limit, we adopt the Kadanoff-Baym method \cite{Baym1961} in this paper. 
The method is valid when the external momentum is at the sub-horizon scale as a particle description is valid. 
By using it, we can systematically obtain the on-shell term and the off-shell term. 
So it is clearly visible when they are degenerate. 
We show that the IR singularities appearing at the on-shell limit cancel out between real and virtual processes also in a matter system with gravity. 

It is an important task to identify observables in dS space. 
There is no analogue of S matrix of QCD here since we cannot observe super-horizon modes. 
We can still measure the couplings of the microscopic theory such as the standard model of particle physics in dS space. 
In fact our Universe is evolving toward such a situation. 
Therefore they are certainly observables in dS space. 
It is an nontrivial issue whether soft graviton effects preserve Lorentz invariance at sub-horizon scale. 
Our investigation addresses such a fundamental question. 

The organization of this paper is as follows. 
In Section \ref{GP}, we quantize the gravitational field on the dS background. 
We make a brief review of the Kadanoff-Baym approach in Section \ref{KB}. 
The subsequent two sections constitute the main results of this paper where we investigate interaction effects through the collision term. 
In our investigation, we focus on the local contribution in Section \ref{cL} and the non-local contribution in Section \ref{cNL}. 
We find that the local term contributes to the dS symmetry breaking while the non-local term does not. 
Furthermore we confirm that soft gravitational effects induced by the local terms preserve the effective Lorentz invariance. 
In Section \ref{Para}, we show that the results obtained in the previous sections do not depend on the parametrization of the metric. 
In Section \ref{Gauge}, we investigate the gauge dependence of soft gravitational effects.  
We conclude with discussions in Section \ref{Conclusion}. 

\section{Gravitational propagators in dS space}\label{GP}
\setcounter{equation}{0}

In this section, we review gravitational propagators in dS space. 
In the Poincar\'{e} coordinate, the metric of dS space is written as 
\begin{align}
ds^2&=-dt^2+a^2(t)d{\bf x}^2 \\
&=a^2(\tau)(-d\tau^2+d{\bf x}^2), \notag
\end{align}
\begin{align}
a=e^{Ht}=-\frac{1}{H\tau}, 
\end{align}
where the Hubble parameter $H$ is constant and the dimension of dS space is taken as $D=4$. 
The conformal time $\tau$ runs in the range: $-\infty<\tau<0$. 
After a sufficient exponential expansion, the dS space is well described locally by the above metric irrespective of the spatial topology. 

In investigating gravitational fluctuations, we primarily use the following parametrization of the metric: 
\begin{align}
g_{\mu\nu}=\Omega^2(x)\tilde{g}_{\mu\nu},\hspace{1em}\Omega(x)=a(\tau)e^{\kappa w(x)}, 
\label{para1}\end{align}
\begin{align}
\det \tilde{g}_{\mu\nu}=-1,\hspace{1em}\tilde{g}_{\mu\nu}=\eta_{\mu\rho}(e^{\kappa h(x)})^{\rho}_{~\nu}, 
\label{para2}\end{align}
where $\kappa$ is defined by the Newton's constant $G$ as $\kappa^2=16\pi G$. 
To satisfy (\ref{para2}), $h_{\mu\nu}$ is taken to be traceless
\begin{align}
h_\mu^{\ \mu}=0. 
\label{para3}\end{align}
In this parametrization, the scalar density and the Ricci scalar are written as 
\begin{align}
\sqrt{-g}=\Omega^4,\hspace{1em}
R=\Omega^{-2}\tilde{R}-6\Omega^{-3}\tilde{g}^{\mu\nu}\nabla_\mu\partial_\nu\Omega, 
\label{R1}\end{align}
where $\tilde{R}$ is the Ricci scalar constructed from $\tilde{g}_{\mu\nu}$
\begin{align}
\tilde{R}=-\partial_\mu\partial_\nu\tilde{g}^{\mu\nu}
-\frac{1}{4}\tilde{g}^{\mu\nu}\tilde{g}^{\rho\sigma}\tilde{g}^{\alpha\beta}\partial_\mu\tilde{g}_{\rho\alpha}\partial_\nu\tilde{g}_{\sigma\beta}
+\frac{1}{2}\tilde{g}^{\mu\nu}\tilde{g}^{\rho\sigma}\tilde{g}^{\alpha\beta}\partial_\mu\tilde{g}_{\sigma\alpha}\partial_\rho\tilde{g}_{\nu\beta}. 
\label{R2}\end{align}
By using the partial integration, the Lagrangian density for the Einstein gravity is written as
\begin{align}
\mathcal{L}_\text{Gravity}
=\frac{1}{\kappa^2}\sqrt{-g}\big[R-2\Lambda\big]
=\frac{1}{\kappa^2}\big[\Omega^2\tilde{R}+6\tilde{g}^{\mu\nu}\partial_\mu\Omega\partial_\nu\Omega-6H^2\Omega^4\big], 
 \label{Gravity}\end{align}
where $\Lambda=3H^2$. 

In order to fix the gauge with respect to general coordinate invariance, we adopt the following gauge fixing term \cite{Tsamis1992}: 
\begin{align}
\mathcal{L}_\text{GF}&=-\frac{1}{2}a^{2}F_\mu F^\mu, \label{GF}\\
F_\mu&=\partial_\rho h_\mu^{\ \rho}-2\partial_\mu w+2h_\mu^{\ \rho}\partial_\rho\log a+4w\partial_\mu\log a. \notag
\end{align}
Note that in this paper, the Lagrangian density is defined including $\sqrt{-g}$ 
and the Lorentz indexes are raised and lowered by $\eta^{\mu\nu}$ and $\eta_{\mu\nu}$ respectively. 
The corresponding ghost term at the quadratic level is given by
\begin{align}
\mathcal{L}_\text{ghost}=&-a^2\partial^\nu\bar{b}^\mu 
\big\{{\eta}_{\mu\rho}\partial_\nu+\eta_{\nu\rho}\partial_\mu+2\eta_{\mu\nu}\partial_\rho(\log a)\big\}b^\rho\label{ghost}\\
&+\partial_\mu(a^2\bar{b}^\mu)
\big\{\partial_\nu+4\partial_\nu(\log a)\big\}b^\nu, \notag
\end{align}
where $b^\mu$ is the ghost field and $\bar{b}^\mu$ is the anti-ghost field. 
From (\ref{Gravity})-(\ref{ghost}), the quadratic part of the total gravitational Lagrangian density is
\begin{align}
\mathcal{L}_\text{quadratic}=a^4&\big[\ \frac{1}{2}a^{-2}\partial_\mu X\partial^\mu X-\frac{1}{4}a^{-2}\partial_\mu\tilde{h}^{ij}\partial^\mu\tilde{h}^{ij}
-a^{-2}\partial_\mu\bar{b}^i\partial^\mu b^i\label{quadratic}\\
&+\frac{1}{2}a^{-2}\partial_\mu h^{0i}\partial^\mu h^{0i}+H^2h^{0i}h^{0i}
-\frac{1}{2}a^{-2}\partial_\mu Y\partial^\mu Y-H^2Y^2\notag\\
&+a^{-2}\partial_\mu\bar{b}^0\partial^\mu b^0+2H^2\bar{b}^0b^0\big]. \notag
\end{align}
Here we have decomposed $h^i_{\ j},\ i,j=1, \cdots, 3$ into the trace and traceless part
\begin{align}
h^{ij}=\tilde{h}^{ij}+\frac{1}{3}h^{kk}\delta^{ij}=\tilde{h}^{ij}+\frac{1}{3}h^{00}\delta^{ij}.
\end{align} 
The action has been diagonalized by the following linear combination 
\begin{align}
X=2\sqrt{3}w-\frac{1}{\sqrt{3}}h^{00},\hspace{1em}Y=h^{00}-2w. 
\label{diagonalize}\end{align}
The quadratic action (\ref{quadratic}) contains two types of fields, 
massless and minimally coupled fields: $X,\tilde{h}^{ij},b^i,\bar{b}^i$ 
and massless conformally coupled fields: $h^{0i},b^0,\bar{b}^0$, $Y$. 
We list the corresponding propagators as follows 
\begin{align}
\langle X(x)X(x')\rangle&=-\langle\varphi(x)\varphi(x')\rangle, \label{minimally}\\
\langle\tilde{h}^{ij}(x)\tilde{h}^{kl}(x')\rangle&=(\delta^{ik}\delta^{jl}+\delta^{il}\delta^{jk}-\frac{2}{3}\delta^{ij}\delta^{kl})\langle\varphi(x)\varphi(x')\rangle, \notag\\
\langle b^i(x)\bar{b}^j(x')\rangle&=\delta^{ij}\langle\varphi(x)\varphi(x')\rangle, \notag
\end{align}
\begin{align}
\langle h^{0i}(x)h^{0j}(x')\rangle&=-\delta^{ij}\langle\phi(x)\phi(x')\rangle, \label{conformally}\\
\langle Y(x)Y(x')\rangle&=\langle\phi(x)\phi(x')\rangle, \notag\\
\langle b^0(x)\bar{b}^0(x')\rangle&=-\langle\phi(x)\phi(x')\rangle. \notag
\end{align}
We should mention that the spatial traceless part $\tilde{h}^{ij}$ consists of not only the tensor mode but the vector and scalar modes: 
\begin{align}
\langle\tilde{h}_{ij}(x)\tilde{h}_{kl}(x')\rangle=\langle\tilde{h}^T_{ij}(x)\tilde{h}^T_{kl}(x')\rangle+\langle\tilde{h}^V_{ij}(x)\tilde{h}^V_{kl}(x')\rangle+\langle\tilde{h}^S_{ij}(x)\tilde{h}^S_{kl}(x')\rangle, 
\end{align}
\begin{align}
\langle\tilde{h}^T_{ij}(x)\tilde{h}^T_{kl}(x')\rangle&=(\bar{\delta}_{ik}\bar{\delta}_{jl}+\bar{\delta}_{il}\bar{\delta}_{jk}-\bar{\delta}_{ij}\bar{\delta}_{kl})\langle\varphi(x)\varphi(x')\rangle, \\
\langle\tilde{h}^V_{ij}(x)\tilde{h}^V_{kl}(x')\rangle&=(\bar{\delta}_{ik}\frac{\partial_j\partial_l}{\partial_m^2}+\bar{\delta}_{jl}\frac{\partial_i\partial_k}{\partial_m^2}
+\bar{\delta}_{il}\frac{\partial_j\partial_k}{\partial_m^2}+\bar{\delta}_{jk}\frac{\partial_i\partial_l}{\partial_m^2})\langle\varphi(x)\varphi(x')\rangle, \notag\\
\langle\tilde{h}^S_{ij}(x)\tilde{h}^S_{kl}(x')\rangle&=3(\frac{\partial_i\partial_j}{\partial_m^2}-\frac{1}{3}\delta_{ij})(\frac{\partial_k\partial_l}{\partial_m^2}-\frac{1}{3}\delta_{kl})\langle\varphi(x)\varphi(x')\rangle, \notag
\end{align} 
where $\bar{\delta}_{ij}$ denotes the projection to the transverse part: $\bar{\delta}_{ij}\equiv \delta_{ij}-\partial_i\partial_j/\partial_m^2$. 
The degeneracy of these modes is a specific property in the gauge condition (\ref{GF}). 
In Section \ref{Gauge}, we investigate the gravitational propagator in a more general gauge condition.  

In (\ref{minimally}) and (\ref{conformally}), $\varphi$ denotes a massless and minimally coupled scalar field and $\phi$ denotes a massless conformally coupled scalar field. 
The corresponding wave functions $\varphi_{\bf p}(x)$, $\phi_{\bf p}(x)$ are given by
\begin{align}
\varphi_{\bf p}(x)=\frac{H\tau}{\sqrt{2p}}(1-i\frac{1}{p\tau})e^{-ip\tau+i{\bf p}\cdot{\bf x}},  
\label{minimallywf}\end{align}
\begin{align}
\phi_{\bf p}(x)=\frac{H\tau}{\sqrt{2p}}e^{-ip\tau+i{\bf p}\cdot{\bf x}}. 
\label{conformallywf}\end{align}
The massless and conformally coupled field is locally equal to that in Minkowski space up to the scale factor
\begin{align}
\langle\phi(x)\phi(x')\rangle&=\frac{H^2}{4\pi^2}\frac{1}{y},  
\label{conformally0}\end{align}
where $y$ is the square of distance which preserves the dS symmetry
\begin{align}
y=\Delta x^2/\tau\tau',\hspace{1em}\Delta x^2=-(\tau-\tau')^2+({\bf x}-{\bf x}')^2. 
\end{align}

On the other hand, the massless and minimally coupled field has a specific property to dS space. 
At the super-horizon scale as physical momentum: $P\equiv p/a(\tau)\ll H$, the wave function (\ref{minimallywf}) behaves as
\begin{align}
\varphi_{\bf p}(x)\sim\frac{H}{\sqrt{2p^3}}e^{i{\bf p}\cdot{\bf x}}. 
\end{align}
The IR behavior indicates that the corresponding propagator has a logarithmic divergence from the IR contributions in the infinite volume limit. 
To regularize the IR divergence, we introduce an IR cut-off $1/L_i$ which fixes the minimum value of the comoving momentum. 
Physically speaking, $L_i$ is recognized as an initial size of universe when the exponential expansion starts. 
Due to the commutation relation, it is equivalent to set the initial time as
\begin{align}
a_i=-1/H\tau_i=1/H L_i. 
\end{align}
With this prescription, the propagator for a massless and minimally coupled field is given by
\begin{align}
\langle\varphi(x)\varphi(x')\rangle&=\frac{H^2}{4\pi^2}\big\{\frac{1}{y}-\frac{1}{2}\log y+\frac{1}{2}\log \big(a(\tau)a(\tau')/a_i^2\big)+1-\gamma\big\}, 
\label{minimally0}\end{align} 
where $\gamma$ is Euler's constant.  
The existence of  the logarithmic term: $\log \big(a(\tau)a(\tau')/a_i^2\big)$ implies the breakdown of the dS symmetry. 
In particular, it breaks the scale invariance
\begin{align}
\tau\to C\tau,\hspace{1em}x^i\to Cx^i. 
\label{scaling}\end{align}
To explain what causes the dS symmetry breaking,  
we recall that the minimum value of the physical momentum is $1/a(\tau )L_i$ as the wavelength is stretched by cosmic expansion. 
That is, more degrees of freedom accumulate at the super-horizon scale with cosmic evolution. 
Due to this increase, the propagator acquires the growing time dependence which spoils the dS symmetry. 

As there is explicit time dependence in the propagator, 
physical quantities can acquire time dependence through the quantum loop corrections. 
We call them the quantum IR effects in dS space. 
In order to clearly separate the minimally coupled modes and the conformally coupled modes, the gravitational propagator is written as
\begin{align}
\langle h^{\mu\nu}(x)h^{\rho\sigma}(x')\rangle=P^{\mu\nu\rho\sigma}\langle\varphi(x)\varphi(x')\rangle+Q^{\mu\nu\rho\sigma}\langle\phi(x)\phi(x')\rangle, \label{gravityp}
\end{align}
\begin{align}
P^{\mu\nu\rho\sigma}=&-\frac{3}{4}\delta^\mu_{\ 0}\delta^\nu_{\ 0}\delta^\rho_{\ 0}\delta^\sigma_{\ 0}
-\frac{1}{4}(\delta^\mu_{\ 0}\delta^\nu_{\ 0}\check{\eta}^{\rho\sigma}+\delta^\rho_{\ 0}\delta^\sigma_{\ 0}\check{\eta}^{\mu\nu}) \label{P}\\
&+(\check{\eta}^{\mu\rho}\check{\eta}^{\nu\sigma}+\check{\eta}^{\mu\sigma}\check{\eta}^{\nu\rho}-\frac{3}{4}\check{\eta}^{\mu\nu}\check{\eta}^{\rho\sigma}), \notag
\end{align}
\begin{align}
Q^{\mu\nu\rho\sigma}=&+\frac{9}{4}\delta^\mu_{\ 0}\delta^\nu_{\ 0}\delta^\rho_{\ 0}\delta^\sigma_{\ 0}
+\frac{3}{4}(\delta^\mu_{\ 0}\delta^\nu_{\ 0}\check{\eta}^{\rho\sigma}+\delta^\rho_{\ 0}\delta^\sigma_{\ 0}\check{\eta}^{\mu\nu}) \label{Q}\\
&-(\delta^\mu_{\ 0}\delta^\rho_{\ 0}\check{\eta}^{\nu\sigma}+\delta^\mu_{\ 0}\delta^\sigma_{\ 0}\check{\eta}^{\nu\rho}+\delta^\nu_{\ 0}\delta^\rho_{\ 0}\check{\eta}^{\mu\sigma}+\delta^\nu_{\ 0}\delta^\sigma_{\ 0}\check{\eta}^{\mu\rho}) \notag\\
&+\frac{1}{4}\check{\eta}^{\mu\nu}\check{\eta}^{\rho\sigma}. \notag
\end{align}
Here $\check{\eta}^{\mu\nu}$ denotes the projection to the spatial subspace: $\check{\eta}^{\mu\nu}\equiv \eta^{\mu\nu}+\delta^\mu_{\ 0}\delta^\nu_{\ 0}$. 

Before closing this section, we refer to the parametrization dependence of the metric. 
There are other choices in the parametrization of the metric than (\ref{para1})-(\ref{para3}). 
For example, the following parametrization is adopted in \cite{Tsamis1992,Kahya2007,Miao2005}: 
\begin{align}
g_{\mu\nu}=a^2(\tau)(\eta_{\mu\nu}+2\kappa\Phi(x)\eta_{\mu\nu}+\kappa\Psi_{\mu\nu}(x)),\hspace{1em}\Psi_\mu^{\ \mu}=0. 
\label{Wpara}\end{align}
Here we have divided the fluctuation into the trace and traceless part to facilitate the comparison with the parametrization (\ref{para1})-(\ref{para3}). 
The relation of them is given by 
\begin{align}
\kappa w&=\kappa\Phi-\kappa^2\Phi^2-\frac{1}{16}\kappa^2\Psi_{\rho\sigma}\Psi^{\rho\sigma}+\cdots, \label{difference}\\
\kappa h_{\mu\nu}&=\kappa\Psi_{\mu\nu}-2\kappa^2\Phi\Psi_{\mu\nu}-\frac{1}{2}\kappa^2\Psi_\mu^{\ \rho}\Psi_{\rho\nu}+\frac{1}{8}\kappa^2\Psi_{\rho\sigma}\Psi^{\rho\sigma}\eta_{\mu\nu}+\cdots. \notag
\end{align} 
We should note that $w$, $h_{\mu\nu}$ is equal to $\Phi$, $\Psi_{\mu\nu}$ up to the linear order. 
As far as we adopt the same gauge: 
\begin{align}
F_\mu=\partial_\rho \Psi_\mu^{\ \rho}-2\partial_\mu \Phi+2\Psi_\mu^{\ \rho}\partial_\rho\log a+4\Phi\partial_\mu\log a, 
\end{align}
we have only to identify the field components to obtain the propagator in the parametrization (\ref{Wpara}): 
\begin{align}
w\to\Phi,\hspace{1em}h_{\mu\nu}\to\Psi_{\mu\nu}. 
\label{rename}\end{align}
The deference between these two parametrizations emerges in the non-linear order. 

\section{Kadanoff-Baym approach}\label{KB}
\setcounter{equation}{0}

As the main subject of this paper, we investigate soft gravitational effects on a matter system by introducing a Kadanoff-Baym method \cite{Baym1961}. 
The investigation is up to the one-loop level. 
Compared to the previous studies with the effective equation of motion, we can systematically take into account the process with a soft or collinear particle in the approach. 
For simplicity, we adopt a massless and conformally coupled scalar field as a matter field
\begin{align}
S=\int d^4x \sqrt{-g}\big[-\frac{1}{2}g^{\mu\nu}\partial_\mu\phi\partial_\nu\phi-\frac{1}{12}R\phi^2\big]. 
\end{align} 
After the field redefinition, 
\begin{align}
\tilde{\phi}\equiv\Omega\phi,  
\end{align} 
the action is written as
\begin{align}
S=\int d^4x\big[-\frac{1}{2}\tilde{g}^{\mu\nu}\partial_\mu\tilde{\phi}\partial_\nu\tilde{\phi}-\frac{1}{12}\tilde{R}\tilde{\phi}^2\big]. 
\label{action}\end{align}
From (\ref{para2}) and (\ref{R2}), we can read off the relevant interaction vertices: 
\begin{align}
S_\text{int}=&\int d^4x\ \big[-\frac{1}{2}(-h^{\mu\nu}+\frac{1}{2}h^\mu_{\ \rho}h^{\rho\nu})\partial_\mu\tilde{\phi}\partial_\nu\tilde{\phi} \label{int}\\
&-\frac{1}{12}\big\{-\partial_\mu\partial_\nu(-h^{\mu\nu}+\frac{1}{2}h^\mu_{\ \rho}h^{\rho\nu})
-\frac{1}{4}\partial_\mu h_{\rho\alpha}\partial^\mu h^{\rho\alpha}+\frac{1}{2}\partial_\mu h_{\rho\alpha}\partial^\rho h^{\mu\alpha}\big\}
\tilde{\phi}^2\big]. \notag
\end{align}
Here we should note that the ghost, anti-ghost fields do not couple to the matter field directly in the gauge condition (\ref{GF}) 
and so they do not contribute to the matter field dynamics up to the one-loop level. 

At the tree level, the corresponding propagator is equal to that in Minkowski space 
\begin{align}
\langle\tilde{\phi}(x_1)\tilde{\phi}(x_2)\rangle=\int\frac{d^3p}{(2\pi)^3}\ \frac{1}{2p}e^{-ip(\tau_1-\tau_2)+i{\bf p}\cdot({\bf x}_1-{\bf x}_2)}. 
\label{free}\end{align}
To investigate interaction effects in a time dependent background like dS space, 
we need to adopt the Schwinger-Keldysh formalism \cite{Schwinger1961,Keldysh1964}. 
We introduce the Schwinger-Keldysh indices as 
\begin{align}
G^{-+}(x_1,x_2)&\equiv\langle\tilde{\phi}(x_1)\tilde{\phi}(x_2) \rangle, \label{SK}\\
G^{+-}(x_1,x_2)&\equiv\langle\tilde{\phi}(x_2)\tilde{\phi}(x_1) \rangle, \notag\\
G^{++}(x_1,x_2)&\equiv\theta(\tau_1-\tau_2)\langle\tilde{\phi}(x_1)\tilde{\phi}(x_2) \rangle+\theta(\tau_2-\tau_1)\langle\tilde{\phi}(x_2)\tilde{\phi(x_1)} \rangle, \notag\\
G^{--}(x_1,x_2)&\equiv\theta(\tau_2-\tau_1)\langle\tilde{\phi}(x_1)\tilde{\phi}(x_2) \rangle+\theta(\tau_1-\tau_2)\langle\tilde{\phi}(x_2)\tilde{\phi}(x_1) \rangle. \notag
\end{align}
Here the propagator $G$ includes interaction effects in general. 
When we specify the free propagator (\ref{free}), it is denoted by $G_0$. 
We recall that as for the free propagator, the following identities hold 
\begin{align}
&G_0^{-1}\equiv i(\partial_0^2-\partial_i^2), \label{-1}\\
&G_0^{-1}|_{x_1}G_0^{ab}(x_1,x_2)=c^{ab}\delta^{(4)}(x_1-x_2), \notag
\end{align}
where $c^{ab}$ is defined as
\begin{align}
c^{ab}=\begin{pmatrix} 1 & 0 \\ 0 & -1 \end{pmatrix},\hspace{1em}a,b=+,-. 
\end{align}

Up to the one-loop level, the Schwinger-Dyson equation is given by
\begin{align}
G^{-+}(x_1,x_2)=&\ G_0^{-+}(x_1,x_2) \label{SD}\\
&+\int d^4x'd^4x''\ c_{ab}\ G_0^{-a}(x_1,x')\Sigma_\text{4-pt}(x',x'')G_0^{b+}(x'',x_2) \notag\\
&+\int d^4x'd^4x''\ c_{ab}c_{cd}\ G_0^{-a}(x_1,x')\Sigma^{bc}_\text{3-pt}(x',x'')G^{d+}_0(x'',x_2), \notag
\end{align}
where the self-energy $\Sigma$ due to the four-point vertices and the three-point vertices are given by
\begin{align}
\Sigma_\text{4-pt}(x,x')=i\delta^{(4)}(x-x')\times\Big[&\ \frac{1}{2}\kappa^2\partial'_\mu\big\{\langle h^\mu_{\ \rho}(x')h^{\rho\nu}(x')\rangle\partial'_\nu\big\}
+\frac{1}{12}\kappa^2\partial'_\mu\partial'_\nu\langle h^\mu_{\ \rho}(x')h^{\rho\nu}(x')\rangle \label{4-pt}\\
&+\frac{1}{24}\kappa^2\langle \partial'_\mu h_{\rho\alpha}(x')\partial'^\mu h^{\rho\alpha}(x')\rangle
-\frac{1}{12}\kappa^2\langle \partial'_\mu h_{\rho\alpha}(x')\partial'^\rho h^{\mu\alpha}(x')\rangle\Big], \notag
\end{align}
\begin{align}
\Sigma^{-+}_\text{3-pt}(x,x')=
&-\kappa^2\partial_\mu\partial'_\sigma\big\{\langle h^{\mu\nu}(x)h^{\rho\sigma}(x')\rangle\langle\partial_\nu\tilde{\phi}(x)\partial'_\rho\tilde{\phi}(x')\rangle\big\} \label{3-pt}\\
&-\frac{1}{6}\kappa^2\partial_\mu\big\{\langle h^{\mu\nu}(x)\partial'_\rho\partial'_\sigma h^{\rho\sigma}(x')\rangle\langle\partial_\nu\tilde{\phi}(x)\tilde{\phi}(x')\rangle\big\} \notag\\
&-\frac{1}{6}\kappa^2\partial'_\sigma\big\{\langle \partial_\mu\partial_\nu h^{\mu\nu}(x)h^{\rho\sigma}(x')\rangle\langle\tilde{\phi}(x)\partial'_\rho\tilde{\phi}(x')\rangle\big\} \notag\\
&-\frac{1}{36}\kappa^2\langle \partial_\mu\partial_\nu h^{\mu\nu}(x)\partial'_\rho\partial'_\sigma h^{\rho\sigma}(x')\rangle\langle\tilde{\phi}(x)\tilde{\phi}(x')\rangle. \notag
\end{align}
As for the other indices, $\Sigma_\text{3-pt}$ is defined in a similar way to the propagator (\ref{SK}). 

By introducing the retarded and the advanced functions: 
\begin{align}
&F^R(x_1,x_2)\equiv \theta(\tau_1-\tau_2)[F^{-+}(x_1,x_2)-F^{+-}(x_1,x_2)], \hspace{1em}F =G,\Sigma_\text{3-pt}, \\
&F^A(x_1,x_2)\equiv -\theta(\tau_2-\tau_1)[F^{-+}(x_1,x_2)-F^{+-}(x_1,x_2)], \notag
\end{align}
the Schwinger-Dyson equation is written as the following form 
\begin{align}
G^{-+}(x_1,x_2)=&\ G^{-+}_0(x_1,x_2) \label{SD1}\\
&+\int d^4x'd^4x''\ G^R_0(x_1,x')\Sigma_\text{4-pt}(x',x'')G_0^{-+}(x'',x_2) \notag\\
&+\int d^4x'd^4x''\ G^{-+}_0(x_1,x')\Sigma_\text{4-pt}(x',x'')G_0^A(x'',x_2) \notag\\
&+\int d^4x'd^4x''\ G^R_0(x_1,x')\Sigma^R_\text{3-pt}(x',x'')G_0^{-+}(x'',x_2) \notag\\
&+\int d^4x'd^4x''\ G^R_0(x_1,x')\Sigma^{-+}_\text{3-pt}(x',x'')G_0^A(x'',x_2) \notag\\
&+\int d^4x'd^4x''\ G^{-+}_0(x_1,x')\Sigma^A_\text{3-pt}(x',x'')G_0^A(x'',x_2). \notag
\end{align} 
As seen in (\ref{3-pt}), $\Sigma_\text{3-pt}$ contains differential operators.  
They are applied after the step functions are assigned. The prescription corresponds with the $T^*$ product.  

We introduce our principle assumption that the full propagator in dS space is written as the following form: 
\begin{align}
G^{-+}(x_1,x_2)=&\int\frac{d^3p}{(2\pi)^3}\ Z(p,\tau_c)\frac{1}{2p}e^{-ip(\tau_1-\tau_2)+i{\bf p}\cdot({\bf x}_1-{\bf x}_2)} \label{full}\\
&+\int_{\epsilon>p}\frac{d\epsilon d^3p}{(2\pi)^3}\ N(p,\epsilon,\tau_c)\frac{1}{2\epsilon}e^{-i\epsilon(\tau_1-\tau_2)+i{\bf p}\cdot({\bf x}_1-{\bf x}_2)}. \notag
\end{align}
The full propagator depends on the average and the relative time
\begin{align}
\tau_c\equiv\frac{\tau_1+\tau_2}{2},\hspace{1em}\Delta \tau\equiv \tau_1-\tau_2. 
\end{align}
Due to the spatial translational symmetry, we can expand it by spatial plane waves. 
The existence of interactions do not allow the full propagator to consist of on-shell term alone. 
Therefore we have introduced the on-shell part and the off-shell part of the spectral function: $Z$, $N$.  
The on-shell part $Z$ is frequently called the wave function renormalization factor. 
Focusing on the region $|\tau_c|\gg |\Delta\tau|$, we assume that they evolve with the average time. 

In this paper, we investigate time dependent quantum effects which break the dS symmetry
in the two-point function of the conformally coupled scalar field. 
The propagator of the massless and minimally coupled field (\ref{minimally0}) in gravitational
modes may induce such a symmetry breaking through gravitational interaction. 
In fact we have identified such IR logarithms which grow at late times \cite{KitamotoSD}. 
However we also find that IR logarithms appear in the form of the wave function renormalization. 
It can be absorbed into the $Z$ factor if we allow it to be time dependent. 
Such an approximation is valid at sub-horizon scale where the change of $Z$ factor is slow. 
Namely $Z$, $N$ are supposed to be time dependent even when they are expressed by the physical scales as
\begin{align}
Z(P,a(\tau_c)/a_i),\hspace{1em}N(P,E,a(\tau_c)/a_i), 
\end{align}
\begin{align}
P\equiv pH|\tau_c|,\hspace{1em}E\equiv \epsilon H|\tau_c|. 
\label{PQ}\end{align}
To evaluate the time dependence which breaks the dS symmetry, we derive the differential equations for $Z$, $N$ by the Kadanoff-Baym approach. 
Our previous investigations indicate that IR effects do not spoil Lorentz invariance at sub-horizon scale.
In particular the velocity of light is not renormalized by IR logarithms irrespective of the spin.
We remark that it is not the case if we only consider graviton (spatial, traceless, transverse) modes.
Thus our postulate (\ref{full}) is base on the Lorentz invariance at sub-horizon scale.
Lorentz invariance is one of the fundamental principles in microscopic physics.
Its validity can be experimentally tested.
In this paper we indeed find the evidences for the consistency of this assumption.

In (\ref{full}), we set the initial state to be the vacuum state. 
We need to introduce a distribution function when we start with an excited state \cite{Polyakov,Kitamoto2010,Akhmedov}. 
We have found that in $\varphi^3$, $\varphi^4$ theories, the non-local IR effects contribute to the spectrum function prior to the distribution function. 
Furthermore, the non-local IR singularities are canceled between $Z$ and $N$ \cite{Kitamoto2010}. 
The main motivation of this paper is to confirm that such a cancellation takes place in the matter system with gravity.  

Phenomenologically we observe physics at some fixed momentum scales. 
In the subsequent discussions, we suppress the following integration factor by performing the Fourier transformation with respect to the spatial coordinate $\Delta{\bf x}={\bf x}_1-{\bf x}_2$
\begin{align}
\int\frac{d^3p}{(2\pi)^3}\ e^{+i{\bf p}\cdot({\bf x}_1-{\bf x}_2)}. 
\end{align}
In order to investigate physics which are directly observable, 
we set the external momentum to be of the sub-horizon scale:   
\begin{align}
p|\tau_c|\gg 1\ \Leftrightarrow\ P\gg H. 
\end{align}
We should mention that the effective mass is not considered in (\ref{full}). 
The mass term is negligible compared with the kinetic term at such a high external momentum scale.

In order to derive the differential equation of $Z$, $N$, we operate $G_0^{-1}$ on the Schwinger-Dyson equation from the left and the right respectively. 
From (\ref{-1}) and (\ref{SD1}), each identity is given by 
\begin{align}
G_0^{-1}|_{x_1}G^{-+}(x_1,x_2) 
=&\ \int d^4x'\ \Sigma_\text{4-pt}(x_1,x')G^{-+}_0(x',x_2) \label{Bleft}\\
&+\int d^4x'\ \Sigma^R_\text{3-pt}(x_1,x')G^{-+}_0(x',x_2)+\int d^4x'\ \Sigma^{-+}_\text{3-pt}(x_1,x')G^A_0(x',x_2), \notag
\end{align}
\begin{align}
G_0^{-1}|_{x_2}G^{-+}(x_1,x_2) 
=&\ \int d^4x'\ G^{-+}_0(x_1,x')\Sigma_\text{4-pt}(x',x_2) \label{Bright}\\
&+\int d^4x'\ G^{-+}_0(x_1,x')\Sigma^A_\text{3-pt}(x',x_2)+\int d^4x'\ G^R_0(x_1,x')\Sigma^{-+}_\text{3-pt}(x',x_2). \notag
\end{align}
By substituting (\ref{full}), the left-hand sides of them are written as follows after the Fourier transformation
\begin{align}
&\ G_0^{-1}|_{x_1}G^{-+}(x_1,x_2) \label{Dleft}\\
\xrightarrow[\text{F.T.}]{}&\ \big\{\frac{1}{2}\partial_{\tau_c}Z(p,\tau_c)+\frac{i}{8p}\partial_{\tau_c}^2 Z(p,\tau_c)\big\} e^{-ip\Delta\tau} \notag\\
&+\int_p^\infty d\epsilon\ \big\{\frac{1}{2}\partial_{\tau_c}N(p,\epsilon,\tau_c)+\frac{i}{8\epsilon}\partial_{\tau_c}^2 N(p,\epsilon,\tau_c)-i\frac{\epsilon^2-p^2}{2\epsilon}N(p,\epsilon,\tau_c)\big\} e^{-i\epsilon\Delta\tau}, \notag
\end{align}
\begin{align}
&\ G_0^{-1}|_{x_2}G^{-+}(x_1,x_2) \label{Dright}\\
\xrightarrow[\text{F.T.}]{}&\ \big\{-\frac{1}{2}\partial_{\tau_c}Z(p,\tau_c)+\frac{i}{8p}\partial_{\tau_c}^2 Z(p,\tau_c)\big\} e^{-ip\Delta\tau} \notag\\
&+\int_p^\infty d\epsilon\ \big\{-\frac{1}{2}\partial_{\tau_c}N(p,\epsilon,\tau_c)+\frac{i}{8\epsilon}\partial_{\tau_c}^2 N(p,\epsilon,\tau_c)-i\frac{\epsilon^2-p^2}{2\epsilon}N(p,\epsilon,\tau_c)\big\} e^{-i\epsilon\Delta\tau}. \notag
\end{align}
Here we have used the free equation of motion
\begin{align}
\partial^2\tilde{\phi}_{\bf p}(x)=0,\hspace{1em}\tilde{\phi}_{\bf p}(x)\equiv a(\tau)\phi_{\bf p}(x). 
\label{Lorentz}\end{align}
To compile them into a simple differential equation, we consider the difference between (\ref{Bleft}) and (\ref{Bright}): 
 \begin{align}
&\ G_0^{-1}|_{x_1}G^{-+}(x_1,x_2)-G_0^{-1}|_{x_2}G^{-+}(x_1,x_2) \label{B0}\\
=&\ \int d^4x'\ \Sigma_\text{4-pt}(x_1,x')G^{-+}_0(x',x_2) \notag\\
&-\int d^4x'\ G^{-+}_0(x_1,x')\Sigma_\text{4-pt}(x',x_2) \notag\\
&+\int d^4x'\ \Sigma^R_\text{3-pt}(x_1,x')G^{-+}_0(x',x_2)+\int d^4x'\ \Sigma^{-+}_\text{3-pt}(x_1,x')G^A_0(x',x_2) \notag\\
&-\int d^4x'\ G^{-+}_0(x_1,x')\Sigma^A_\text{3-pt}(x',x_2)-\int d^4x'\ G^R_0(x_1,x')\Sigma^{-+}_\text{3-pt}(x',x_2). \notag
\end{align}
From (\ref{Dleft}) and (\ref{Dright}), the left-hand side of (\ref{B0}) is given by 
\begin{align}
&\ G_0^{-1}|_{x_1}G^{-+}(x_1,x_2)-G_0^{-1}|_{x_2}G^{-+}(x_1,x_2) \label{D0}\\
\xrightarrow[\text{F.T.}]{}&\ \partial_{\tau_c}Z(p,\tau_c) e^{-ip\Delta\tau}+\int_p^\infty d\epsilon\ \partial_{\tau_c} N(p,\epsilon,\tau_c) e^{-i\epsilon\Delta\tau}. \notag
\end{align}

We can investigate interaction effects through the right-hand side of (\ref{B0}). Thus we call it the collision term. 
Depending on whether the indices $R$, $A$ are assigned to $\Sigma_\text{3-pt}$ or $G$, the latter part of the collision term is separated into the two parts: 
\begin{align}
\int d^4x'\ \Sigma^R_\text{3-pt}(x_1,x')G^{-+}_0(x',x_2)&\xrightarrow[\text{F.T.}]{} e^{-ip\Delta\tau}\times\cdots, \label{on}\\
-\int d^4x'\ G^{-+}_0(x_1,x')\Sigma^A_\text{3-pt}(x',x_2)&\xrightarrow[\text{F.T.}]{} e^{-ip\Delta\tau}\times\cdots, \notag
\end{align}
\begin{align}
\int d^4x'\ \Sigma^{-+}_\text{3-pt}(x_1,x')G^A_0(x',x_2)&\xrightarrow[\text{F.T.}]{} \int^\infty_p d\epsilon\ e^{-i\epsilon\Delta\tau}\times\cdots, \label{off}\\
-\int d^4x'\ G^R_0(x_1,x')\Sigma^{-+}_\text{3-pt}(x',x_2)&\xrightarrow[\text{F.T.}]{} \int^\infty_p d\epsilon\ e^{-i\epsilon\Delta\tau}\times\cdots. \notag
\end{align}
In terms of the characteristic frequency, we call (\ref{on}) the on-shell terms and (\ref{off}) the off-shell terms. 
Obviously the integrals including $\Sigma_\text{4-pt}$ are the on-shell terms.  
Our aim is to evaluate the wave function renormalization factor $Z$ up to $\mathcal{O}(\log a(\tau_c))$. 
Since $Z$ is differentiated in the left-hand side, we need to evaluate the collision term up to $\mathcal{O}(1/p|\tau_c|)$. 
We expand the collision term by the power series in $1/p|\tau_c|$ type factors which can be justified well inside the cosmological horizon. 
It is a kind of the derivative expansion of the Moyal product in the Wigner representation. 
In the subsequent sections, we investigate the collision term in details.  

\section{Local contribution in the collision term}\label{cL}
\setcounter{equation}{0}

In this section, we focus on the local contribution in the collision term. 
The local terms of the self-energy are identified as they are proportional to $\delta^{(4)}(x-x')$. 
As seen in (\ref{4-pt}), the contribution from the four-point vertices contains only the local terms at the one-loop level. 
The coefficients of them consist of propagators at the coincident point. 

The propagator at the coincident point has an ultra-violet divergence (UV) in general. 
In contrast to the IR divergence which originates in the scale invariant spectrum, 
we can regularize UV divergences respecting the dS symmetry. 
Let us recall that we focus on time dependent quantum effects which break the dS symmetry. 
As seen in (\ref{conformally0}) and (\ref{minimally0}), 
the propagator of the conformal coupled modes respects the dS symmetry 
while the propagator of the minimally coupled modes breaks the dS symmetry. 
Up to the second derivative, these propagators at the coincident point are as follows  
\begin{align}
&\langle\phi(x)\phi(x)\rangle=\text{(UV const.)}, \label{conformallyc}\\
&\langle\partial_\mu\phi(x)\phi(x)\rangle=0, \notag\\
&\langle\partial_\mu\phi(x)\partial_\nu\phi(x)\rangle=\text{(UV const.)}\times g_{\mu\nu}, \notag
\end{align}
\begin{align}
&\langle\varphi(x)\varphi(x)\rangle=\frac{H^2}{4\pi^2}\log \big(a(\tau)/a_i\big)+\text{(UV const.)}, \label{minimallyc}\\
&\langle\partial_\mu\varphi(x)\varphi(x)\rangle=\frac{H^3}{8\pi^2}a(\tau)\delta_\mu^{\ 0}, \notag\\
&\langle\partial_\mu\varphi(x)\partial_\nu\varphi(x)\rangle=-\frac{3H^4}{32\pi^2}g_{\mu\nu}. \notag
\end{align} 
We should emphasize that the coefficients of the dS symmetry breaking terms (IR logarithms) are UV finite. 
Concerning the internal loop contributions at the one-loop level, we can clearly separate IR contributions from UV contributions in comparison to Hubble scale.  
As far as IR logarithms are concerned, we can thus safely ignore UV contributions and UV divergences altogether. 

From (\ref{gravityp})-(\ref{Q}) and (\ref{conformallyc})-(\ref{minimallyc}), the coefficients of the $\log a(\tau)$ and $a(\tau)$ terms in (\ref{4-pt}) are evaluated as 
\begin{align}
\Sigma_\text{4-pt}(x,x')\simeq i\delta^{(4)}(x-x')\times
\frac{\kappa^2H^2}{4\pi^2}\Big\{\log \big(a(\tau')/a_i\big)\big(\frac{3}{8}{\partial'_0}^2+\frac{13}{8}{\partial'_i}^2\big)+Ha(\tau')\big(\frac{3}{8}\partial'_0\big)\Big\}.  
\label{4-pt_local}\end{align}

In contrast, the three-point vertices contribute to the local and non-local terms as (\ref{3-pt}). 
To extract the local terms from (\ref{3-pt}), 
it is useful to recall that $\delta(\tau-\tau')$ is derived by differentiating $\theta(\tau-\tau')$. 
As seen in (\ref{on}) and (\ref{off}), the step function is associated with the self-energy in the on-shell terms but not in the off-shell terms. 
Putting aside differential operators, the retarded self-energy is written as
\begin{align}
\Sigma^R_\text{3-pt}(x,x')\propto 
&\ \frac{1}{2}\langle \{h^{\mu\nu}(x),h^{\rho\sigma}(x')\}\rangle\times\theta(\tau-\tau')\langle[\tilde{\phi}(x),\tilde{\phi}(x')]\rangle \label{3-pt_R}\\
&+\theta(\tau-\tau')\langle [h^{\mu\nu}(x),h^{\rho\sigma}(x')]\rangle\times\frac{1}{2}\langle\{\tilde{\phi}(x),\tilde{\phi}(x')\}\rangle. \notag
\end{align}
Here $[\ ,\ ]$ denotes the commutator and $\{\ ,\ \}$ denotes the anti-commutator. 
The advanced self-energy is written as a similar form. 
The dS breaking logarithms come from the symmetric propagators of gravitational fields. 
Thus we may focus on $\delta^{(4)}(x-x')$ derived from the retarded propagators of scalar fields 
\begin{align}
\partial_\mu\partial_\nu\big(\theta(\tau-\tau')\langle[\tilde{\phi}(x),\tilde{\phi}(x')]\rangle\big)\big|_\text{local}
=-i\delta^{(4)}(x-x')\delta_\mu^{\ 0}\delta_\nu^{\ 0}, 
\label{local1}\end{align}
\begin{align}
&\ \partial_\mu\partial_\nu\partial_\rho\big(\theta(\tau-\tau')\langle[\tilde{\phi}(x),\tilde{\phi}(x')]\rangle\big)\big|_\text{local}  \label{local2}\\
=&-i\delta^{(4)}(x-x')
\big\{\delta_\mu^{\ 0}\delta_\nu^{\ 0}\partial'_\rho+\delta_\mu^{\ 0}\delta_\rho^{\ 0}\partial'_\nu+\delta_\nu^{\ 0}\delta_\rho^{\ 0}\partial'_\mu 
-2\delta_\mu^{\ 0}\delta_\nu^{\ 0}\delta_\rho^{\ 0}\partial'_0\big\}, \notag
\end{align}
\begin{align}
&\ \partial_\mu\partial_\nu\partial_\rho\partial_\sigma\big(\theta(\tau-\tau')\langle[\tilde{\phi}(x),\tilde{\phi}(x')]\rangle\big)\big|_\text{local} \label{local3}\\
=&-i\delta^{(4)}(x-x') 
\big\{\ 
\delta_\mu^{\ 0}\delta_\nu^{\ 0}\partial'_\rho\partial'_\sigma+\delta_\mu^{\ 0}\delta_\rho^{\ 0}\partial'_\nu\partial'_\sigma+\delta_\mu^{\ 0}\delta_\sigma^{\ 0}\partial'_\nu\partial'_\rho 
+\delta_\nu^{\ 0}\delta_\rho^{\ 0}\partial'_\mu\partial'_\sigma \notag\\
&\hspace{6.2em}+\delta_\nu^{\ 0}\delta_\sigma^{\ 0}\partial'_\mu\partial'_\rho+\delta_\rho^{\ 0}\delta_\sigma^{\ 0}\partial'_\mu\partial'_\nu \notag\\
&\hspace{6.2em}-2(\delta_\mu^{\ 0}\delta_\nu^{\ 0}\delta_\rho^{\ 0}\partial'_\sigma+\delta_\mu^{\ 0}\delta_\nu^{\ 0}\delta_\sigma^{\ 0}\partial'_\rho
+\delta_\mu^{\ 0}\delta_\rho^{\ 0}\delta_\sigma^{\ 0}\partial'_\nu+\delta_\nu^{\ 0}\delta_\rho^{\ 0}\delta_\sigma^{\ 0}\partial'_\mu)\partial'_0 \notag\\
&\hspace{6.2em}+4\delta_\mu^{\ 0}\delta_\nu^{\ 0}\delta_\rho^{\ 0}\delta_\sigma^{\ 0}{\partial'_0}^2
+\delta_\mu^{\ 0}\delta_\nu^{\ 0}\delta_\rho^{\ 0}\delta_\sigma^{\ 0}\partial'^2\big\}. \notag
\end{align}
Here we have used the fact that the propagator of the conformally coupled scalar field depends only on the relative coordinate $x-x'$. 

From (\ref{gravityp})-(\ref{Q}), (\ref{conformallyc})-(\ref{minimallyc}) and (\ref{3-pt_R})-(\ref{local3}), the following local terms are induced from the three-point vertices
\begin{align}
&\ \Sigma^R_\text{3-pt}(x,x')|_\text{local} \label{3-pt_local}\\
=&\ i\delta^{(4)}(x-x')\times
\frac{\kappa^2H^2}{4\pi^2}\Big\{\log \big(a(\tau')/a_i\big)\big(-\frac{3}{4}{\partial'_0}^2-\frac{5}{4}{\partial'_i}^2\big)+Ha(\tau')\big(-\frac{3}{4}\partial'_0\big)\Big\}. \notag
\end{align}
The sum of (\ref{4-pt_local}) and (\ref{3-pt_local}) is evaluated as 
\begin{align}
&\ \Sigma_\text{4-pt}(x,x')+\Sigma^R_\text{3-pt}(x,x')|_\text{local} \label{total1}\\
=&\ i\delta^{(4)}(x-x')\times\frac{\kappa^2H^2}{4\pi^2}\Big\{\log \big(a(\tau')/a_i\big)\cdot\frac{3}{8}\partial'^2+Ha(\tau')\big(-\frac{3}{8}\partial'_0\big)\Big\}. \notag
\end{align}
In a similar way, we can find the following local terms 
\begin{align}
&-\Sigma_\text{4-pt}(x,x')-\Sigma^A_\text{3-pt}(x,x')|_\text{local} \label{total2}\\
=&-i\delta^{(4)}(x-x')\times\frac{\kappa^2H^2}{4\pi^2}\Big\{\log \big(a(\tau')/a_i\big)\cdot\frac{3}{8}\partial'^2+Ha(\tau')\big(-\frac{3}{8}\partial'_0\big)\Big\}. \notag
\end{align}

Here we have assumed that the non-local terms do not contribute to the dS symmetry breaking. 
In the next section, we show how this assumption is justified. 
By substituting (\ref{D0}) and (\ref{total1})-(\ref{total2}) into (\ref{B0}), the differential equation of the wave function renormalization factor is written as  
\begin{align}
\partial_{\tau_c} Z(\tau_c) e^{-ip\Delta\tau}
&=-\frac{3}{16}\frac{\kappa^2H^2}{4\pi^2}\big\{Ha(\tau_1)+Ha(\tau_2)\big\} e^{-ip\Delta\tau} \\
&\simeq-\frac{3}{8}\frac{\kappa^2H^2}{4\pi^2}Ha(\tau_c) e^{-ip\Delta\tau}. \notag
\end{align}
We should emphasize that the $\log a(\tau)$ term vanishes in the first line due to the classical equation of motion (\ref{Lorentz}). 
Here it is crucial that the IR logarithm emerges as an overall factor of $\partial'^2$ in (\ref{total1}), (\ref{total2}). 
As seen in (\ref{action}), the matter action possesses the Lorentz symmetry at the classical level. 
Since we can neglect the derivative of the IR logarithm at the sub-horizon scale: 
\begin{align}
P\gg H\ \Rightarrow\ \log a(\tau)\partial_\mu\gg \partial_\mu(\log a(\tau)), 
\end{align}
the overall IR logarithm indicates that the Lorentz symmetry is effectively respected even after soft gravitational effects are included. 

Furthermore we have extracted the average time dependence as $|\tau_c|\gg |\Delta\tau|$ in the second line. 
The solution of the differential equation is given by
\begin{align}
Z(\tau_c)=1-\frac{3}{8}\frac{\kappa^2H^2}{4\pi^2}\log \big(a(\tau_c)/a_i\big), 
\label{Z}\end{align}
where we have set the initial condition as $Z(\tau_i)=1$. 
After the wave function renormalization: 
\begin{align}
\tilde{\phi}(x)\to Z^\frac{1}{2}(\tau)\tilde{\phi}(x),\hspace{1em}
\langle\tilde{\phi}(x_1)\tilde{\phi}(x_2)\rangle\to Z^{-1}(\tau_c)\langle\tilde{\phi}(x_1)\tilde{\phi}(x_2)\rangle, 
\label{WR}\end{align}
there is no physical effect from soft gravitons in the free field theory (\ref{action}) at the sub-horizon scale. 
It is consistent with the result obtained in \cite{KitamotoSD}. 

We also refer to the fact that as far as we consider the dynamics at the sub-horizon scale, 
the same result (\ref{Z}) is derived in non-conformally coupled scalar field theories with one exception. 
In the minimally coupled case, we need to include the IR logarithm from soft scalar and hard graviton intermediate state
\footnote{This effect  changes the coefficient $3/8$ to $1/2$ in $Z(\tau_c)$}.

\section{Non-local contribution in the collision term}\label{cNL}
\setcounter{equation}{0}

In this section, we investigate the non-local contribution in the collision term. 
Specifically we investigate the integrals which do not contain the derivative of $\theta(\tau-\tau')$. 
To begin with, let us calculate the spatial integration
\begin{align}
\int d^3x'\ \Sigma_\text{3-pt}^{ab}(x_1,x')G_0^{cd}(x',x_2), 
\end{align}
where $(a,b), (c,d)=(\pm,\mp)$. 
The propagator $G_0(x',x_2)$ with each Schwinger-Keldysh index contains the common spatial plane waves
\begin{align}
\int \frac{d^3p}{(2\pi)^32p}\ e^{+i{\bf p}\cdot({\bf x}'-{\bf x}_2)}. 
\end{align}
From (\ref{gravityp})-(\ref{Q}) and (\ref{3-pt}),  a straightforward but cumbersome calculation leads to the following integral 
\begin{align}
&\int d^3x'\ \Sigma^{-+}_\text{3-pt} (x_1,x')\times \int \frac{d^3p}{(2\pi)^32p}\ e^{+i{\bf p}\cdot({\bf x}'-{\bf x}_2)} \label{-+1}\\
\xrightarrow[\text{F.T.}]{}&-\frac{\kappa^2H^2}{2p}\int \frac{d^3p_1d^3p_2}{(2\pi)^62p_12p_2}\ (2\pi)^3\delta^{(3)}({\bf p}_1+{\bf p}_2-{\bf p}) e^{-i\epsilon(\tau_1-\tau')} \notag\\
&\times \Big[\ \big\{-\frac{1}{48}(\epsilon^2-p^2)(37\epsilon^2+11p^2)\frac{1}{p_2^2}+\frac{1}{12}(37\epsilon^3-13\epsilon p^2)\frac{1}{p_2}-\frac{17}{6}\epsilon^2+\epsilon p_2\big\} \notag\\
&\hspace{1.6em}+i(\tau_1-\tau')(\epsilon^2-p^2)\big\{-\frac{1}{48}(37\epsilon^2+11p^2)\frac{1}{p_2}+\frac{3}{4}\epsilon-\frac{1}{6}p_2\big\} \notag\\
&\hspace{1.6em}+\frac{7}{24}\tau_1\tau'(\epsilon^2-p^2)^2\ \Big]. \notag
\end{align}
Here ${\bf p}_1$ and ${\bf p}_2$ are respectively the comoving momenta of the intermediate scalar and gravitational fields. 
Furthermore we have introduced the total energy of intermediate particles as 
\begin{align}
\epsilon\equiv p_1+p_2. 
\end{align}
We should emphasize that the integral (\ref{-+1}) has no IR divergence at $p_2=0\ (\epsilon=p)$. 
If the non-local terms contribute to the dS symmetry breaking, it appears after performing the remaining time integral. 

To facilitate the subsequent discussions, we adopt $\epsilon$ and $p_2$ as the integral variables: 
\begin{align}
\frac{1}{2p}\int \frac{d^3p_1d^3p_2}{(2\pi)^62p_12p_2}\ (2\pi)^3\delta^{(3)}({\bf p}_1+{\bf p}_2-{\bf p}) 
=\frac{1}{32\pi^2p^2}\int^\infty_p d\epsilon\int^\frac{\epsilon+p}{2}_\frac{\epsilon-p}{2}dp_2. 
\end{align}
After performing the integral over $p_2$, the integral (\ref{-+1}) is given by 
\begin{align}
-\frac{\kappa^2H^2}{32\pi^2p^2}\int^\infty_p d\epsilon\ A(p,\epsilon,\tau_1,\tau') e^{-i\epsilon(\tau_1-\tau')}, 
\label{-+2}\end{align}
\begin{align}
A(p,\epsilon,\tau_1,\tau')=\Big[&\ \big\{\frac{1}{12}(37\epsilon^3-13\epsilon p^2)\log \frac{\epsilon+p}{\epsilon-p}-\frac{65}{12}\epsilon^2p-\frac{11}{12}p^3\big\} \label{-+3}\\
&+i(\tau_1-\tau')(\epsilon^2-p^2)\big\{-\frac{1}{48}(37\epsilon^2+11p^2)\log \frac{\epsilon+p}{\epsilon-p}+\frac{2}{3}\epsilon p\big\} \notag\\
&+\frac{7}{24}\tau_1\tau'(\epsilon^2-p^2)^2p\ \Big]. \notag
\end{align}
We also show the integral which contains $\Sigma^{+-}_\text{3-pt}(x_1,x')$
\begin{align}
&\int d^3x'\ \Sigma^{+-}_\text{3-pt} (x_1,x')\times \int \frac{d^3p}{(2\pi)^32p}\ e^{+i{\bf p}\cdot({\bf x}'-{\bf x}_2)} \label{+-}\\
\xrightarrow[\text{F.T.}]{}&-\frac{\kappa^2H^2}{32\pi^2p^2}\int^\infty_p d\epsilon\ A^*(p,\epsilon,\tau_1,\tau')e^{+i\epsilon(\tau_1-\tau')}. \notag
\end{align}

Let us evaluate the following integral in (\ref{B0})
\begin{align}
&\int d^4x'\ \Sigma^R_\text{3-pt}(x_1,x')|_\text{non-local}\ G^{-+}(x',x_2) \label{on1}\\
=&\int d^4x'\ \theta(\tau_1-\tau')\big[\Sigma^{-+}(x_1,x')-\Sigma^{+-}(x_1,x')\big]G^{-+}(x',x_2) \notag\\
\xrightarrow[\text{F.T.}]{}&-\frac{\kappa^2H^2}{32\pi^2p^2}\int^\infty_p d\epsilon\int^{\tau_1}_{-\infty}d\tau'\ A(p,\epsilon,\tau_1,\tau')e^{-i\epsilon(\tau_1-\tau')-ip(\tau'-\tau_2)} \notag\\
&+\frac{\kappa^2H^2}{32\pi^2p^2}\int^\infty_p d\epsilon\int^{\tau_1}_{-\infty}d\tau'\ A^*(p,\epsilon,\tau_1,\tau')e^{+i\epsilon(\tau_1-\tau')-ip(\tau'-\tau_2)}. \notag 
\end{align}
It should be noted that we have retained the integral in the second line which does not contain derivatives of $\theta(\tau_1-\tau')$. 
To evaluate the integrals over time, we use the following identities
\begin{align}
\int^{\tau_1}_{-\infty} d\tau'\ e^{\mp i\epsilon(\tau_1-\tau')-ip(\tau'-\tau_2)}&=\frac{1}{i(\pm \epsilon-p)}e^{-ip(\tau_1-\tau_2)}, \\
\int^{\tau_1}_{-\infty} d\tau'\ \tau' e^{\mp i\epsilon(\tau_1-\tau')-ip(\tau'-\tau_2)}&=\big\{\frac{\tau_1}{i(\pm \epsilon-p)}+\frac{1}{(\pm \epsilon-p)^2}\big\}e^{-ip(\tau_1-\tau_2)}, \notag
\end{align}
where the order of double-sign corresponds. These identities indicate that the integral (\ref{on1}) contributes to the on-shell terms. 
After the time integration, the integral (\ref{on1}) is given by
\begin{align}
+i\frac{\kappa^2H^2}{32\pi^2p^2}&e^{-ip(\tau_1-\tau_2)}\int^\infty_p d\epsilon \label{on2}\\
&\times\Big[\ \frac{1}{\epsilon-p}(-\frac{19}{4}\epsilon^2p+\frac{2}{3}\epsilon p^2-\frac{11}{12}p^3)
+(\frac{37}{16}\epsilon^2+\frac{37}{24}\epsilon p+\frac{11}{48}p^2)\log\frac{\epsilon+p}{\epsilon-p} \notag\\
&\hspace{1.8em}+i\frac{7}{24}\tau_1(\epsilon+p)^2p+\frac{7}{24}\tau_1^2(\epsilon-p)(\epsilon+p)^2p\ \Big] \notag\\
+i\frac{\kappa^2H^2}{32\pi^2p^2}&e^{-ip(\tau_1-\tau_2)}\int^\infty_p d\epsilon \notag\\
&\times\Big[\ \frac{1}{\epsilon+p}(-\frac{19}{4}\epsilon^2p-\frac{2}{3}\epsilon p^2-\frac{11}{12}p^3)
+(\frac{37}{16}\epsilon^2-\frac{37}{24}\epsilon p+\frac{11}{48}p^2)\log\frac{\epsilon+p}{\epsilon-p} \notag\\
&\hspace{1.8em}-i\frac{7}{24}\tau_1(\epsilon-p)^2p+\frac{7}{24}\tau_1^2(\epsilon+p)(\epsilon-p)^2p\ \Big]. \notag 
\end{align}
There is an IR divergence at $\epsilon=p$ where a soft or collinear particle is present. 
We should note that the IR divergence originates in the following integrand in (\ref{on1})
\begin{align}
\Sigma^{-+}_\text{3-pt}(x_1,x')G^{-+}(x',x_2). 
\label{common}\end{align} 
Our goal is to identify possible IR logarithms in (\ref{on2}). 
We discard UV power divergent terms as they do not induce logarithms. 
Their IR contribution can be safely neglected. 
We estimate the logarithmically singular part of  (\ref{on2}) as 
\begin{align}
-ip\frac{5\kappa^2H^2}{32\pi^2}e^{-ip(\tau_1-\tau_2)}\int^\infty_p d\epsilon\ 
\frac{1}{\epsilon-p}. 
\label{on3}\end{align}
Since the choice of the initial time corresponds to the IR cut-off of $\epsilon-p$ as 
\begin{align}
\int_{\tau_i} d\tau'\ \to\ \int_{p+|1/\tau_i|}d\epsilon, 
\label{initial}\end{align}
the IR behavior seems to contributes to the dS symmetry breaking. 
However it is well known that in flat space, the IR singularities in the process with a soft or collinear particle are canceled 
after summing over degenerate states between real and virtual processes \cite{Kinoshita1962,Lee1964}. 
This is a universal phenomenon in any unitary theory as we have shown to be the case with $\varphi^3$, $\varphi^4$ theories in dS space \cite{Kitamoto2010}.
We can argue that the analogous cancellation holds in the matter field theory interacting with soft gravitons. 

To confirm the cancellation, let us evaluate another integral in (\ref{B0})
\begin{align}
&\int d^4x'\ \Sigma^{-+}_\text{3-pt}(x_1,x')G^A(x',x_2) \label{off1}\\
=&-\int d^4x'\ \theta(\tau_2-\tau')\Sigma^{-+}_\text{3-pt}(x_1,x')\big[G^{-+}(x',x_2)-G^{+-}(x',x_2)\big] \notag\\
\xrightarrow[\text{F.T.}]{}&+\frac{\kappa^2H^2}{32\pi^2p^2}\int^\infty_p d\epsilon\int^{\tau_2}_{-\infty}d\tau'\ A(p,\epsilon,\tau_1,\tau')e^{-i\epsilon(\tau_1-\tau')-ip(\tau'-\tau_2)} \notag\\
&-\frac{\kappa^2H^2}{32\pi^2p^2}\int^\infty_p d\epsilon\int^{\tau_2}_{-\infty}d\tau'\ A(p,\epsilon,\tau_1,\tau')e^{-i\epsilon(\tau_1-\tau')+ip(\tau'-\tau_2)}. \notag 
\end{align}
From the identities, 
\begin{align}
\int^{\tau_2}_{-\infty} d\tau'\ e^{-i\epsilon(\tau_1-\tau')\mp ip(\tau'-\tau_2)}&=\frac{1}{i(\epsilon \mp p)}e^{-i\epsilon(\tau_1-\tau_2)}, \\
\int^{\tau_2}_{-\infty} d\tau'\ \tau' e^{-i\epsilon(\tau_1-\tau')\mp ip(\tau'-\tau_2)}&=\big\{\frac{\tau_2}{i(\epsilon \mp p)}+\frac{1}{(\epsilon \mp p)^2}\big\}e^{-i\epsilon(\tau_1-\tau_2)}, \notag
\end{align}
the integral contributes to the off-shell terms. 
After the time integration, the integral (\ref{off1}) is given by 
\begin{align}
-i\frac{\kappa^2H^2}{32\pi^2p^2}&\int^\infty_p d\epsilon\ e^{-i\epsilon(\tau_1-\tau_2)} \label{off2}\\
&\times\Big[\ \frac{1}{\epsilon-p}(-\frac{19}{4}\epsilon^2p+\frac{2}{3}\epsilon p^2-\frac{11}{12}p^3)
+(\frac{37}{16}\epsilon^2+\frac{37}{24}\epsilon p+\frac{11}{48}p^2)\log\frac{\epsilon+p}{\epsilon-p} \notag\\
&\hspace{1.8em}+i(\tau_1-\tau_2)(\epsilon+p)\big\{\frac{2}{3}\epsilon p-\frac{1}{48}(37\epsilon^2+11p^2)\log\frac{\epsilon+p}{\epsilon-p}\big\} \notag\\
&\hspace{1.8em}+i\frac{7}{24}\tau_1(\epsilon+p)^2p+\frac{7}{24}\tau_1\tau_2(\epsilon-p)(\epsilon+p)^2p\ \Big] \notag\\
+i\frac{\kappa^2H^2}{32\pi^2p^2}&\int^\infty_p d\epsilon\ e^{-i\epsilon(\tau_1-\tau_2)} \notag\\
&\times\Big[\ \frac{1}{\epsilon+p}(-\frac{19}{4}\epsilon^2p-\frac{2}{3}\epsilon p^2-\frac{11}{12}p^3)
+(\frac{37}{16}\epsilon^2-\frac{37}{24}\epsilon p+\frac{11}{48}p^2)\log\frac{\epsilon+p}{\epsilon-p} \notag\\
&\hspace{1.8em}+i(\tau_1-\tau_2)(\epsilon-p)\big\{\frac{2}{3}\epsilon p-\frac{1}{48}(37\epsilon^2+11p^2)\log\frac{\epsilon+p}{\epsilon-p}\big\} \notag\\
&\hspace{1.8em}+i\frac{7}{24}\tau_1(\epsilon-p)^2p+\frac{7}{24}\tau_1\tau_2(\epsilon+p)(\epsilon-p)^2p\ \Big]. \notag 
\end{align}
The off-shell term also has an IR divergence at $\epsilon=p$ which originates in the common integrand (\ref{common}) with the on-shell term (\ref{on1}). 
The logarithmic IR singularity of (\ref{off2}) is evaluated as
\begin{align}
+ip\frac{5\kappa^2H^2}{32\pi^2}\int^\infty_p d\epsilon\ e^{-i\epsilon(\tau_1-\tau_2)}\frac{1}{\epsilon-p}. 
\label{off3}\end{align}
The differences between (\ref{on3}) and (\ref{off3}) turns out to be the relative opposite sign and their frequencies $p$ and $\epsilon$. 

Physically speaking, any experiment has a finite energy resolution of observation $\Delta \epsilon$. 
We may divide the integral region of the off-shell term as
\begin{align}
\int^\infty_pd\epsilon\ e^{-i\epsilon(\tau_1-\tau_2)}
=\int^\infty_{p+\Delta\epsilon}d\epsilon\ e^{-i\epsilon(\tau_1-\tau_2)}+\int^{p+\Delta\epsilon}_pd\epsilon\ e^{-i\epsilon(\tau_1-\tau_2)}. 
\end{align}
Within the energy resolution, we cannot distinguish the off-shell term from the on-shell term 
\begin{align}
\int^{p+\Delta\epsilon}_pd\epsilon\ e^{-i\epsilon(\tau_1-\tau_2)}\sim e^{-ip(\tau_1-\tau_2)}\int^{p+\Delta\epsilon}_pd\epsilon. 
\end{align}
Thus we need to redefine the on-shell term by transferring the contribution of the off-shell term within the energy resolution $p<\epsilon<p+\Delta\epsilon$: 
\begin{align}
&-ip\frac{5\kappa^2H^2}{32\pi^2}e^{-ip(\tau_1-\tau_2)}\big\{\int^\infty_p d\epsilon-\int^{p+\Delta\epsilon}_p d\epsilon\big\}\ 
\frac{1}{\epsilon-p} \label{on4}\\
=&-ip\frac{5\kappa^2H^2}{32\pi^2}e^{-ip(\tau_1-\tau_2)}\int^\infty_{p+\Delta\epsilon} d\epsilon\ 
\frac{1}{\epsilon-p}. \notag
\end{align}
The remaining contribution is the well-defined off-shell term: 
\begin{align}
+ip\frac{5\kappa^2H^2}{32\pi^2}\int^\infty_{p+\Delta\epsilon} d\epsilon\ e^{-i\epsilon(\tau_1-\tau_2)}\ 
\frac{1}{\epsilon-p}. 
\label{off4}\end{align}
We have found that there is no IR divergence after the redefinition. 
Since the energy resolution of observation is at the sub-horizon scale $\Delta\epsilon\sim{1/ \Delta\tau}\gg|1/\tau_c|>|1/\tau_i|$, 
the IR cut-off is given by not the inverse of the initial time $|1/\tau_i|$ but the energy resolution $\Delta\epsilon$. 

Furthermore we can show the mechanism how the cancellation takes place. 
As seen in (\ref{initial}), the dS symmetry breaking is expressed as the dependence of the initial time. 
The contribution from the negatively large conformal time region is dominant only when the frequency vanishes. 
The zero frequency process is contained in the common integrand ({\ref{common}}) in (\ref{on1}) and (\ref{off1}). 
Then the initial time dependence is canceled as follows:
\begin{align}
&\ \int d^4x'\ \Sigma^R_\text{3-pt}(x_1,x')|_\text{non-local}\ G^{-+}(x',x_2)+\int d^4x'\ \Sigma^{-+}_\text{3-pt}(x_1,x')G^A(x',x_2) \label{another1}\\
\simeq&\ \big\{\int^{\tau_1}_{\tau_i}d\tau'-\int^{\tau_2}_{\tau_i}d\tau'\big\}\int d^3x'\ \Sigma^{-+}_\text{3-pt}(x_1,x')G^{-+}(x',x_2). \notag
\end{align} 
The cancellation holds between the remaining two integrals in (\ref{B0}):
\begin{align}
&-\int d^4x'\ G^{-+}(x_1,x')\Sigma^A_\text{3-pt}(x',x_2)|_\text{non-local}-\int d^4x'\ G^R(x_1,x')\Sigma^{-+}_\text{3-pt}(x',x_2) \label{another2}\\
\simeq&\ \big\{\int^{\tau_2}_{\tau_i}d\tau'-\int^{\tau_1}_{\tau_i}d\tau'\big\}\int d^3x'\ G^{-+}(x_1,x')\Sigma^{-+}_\text{3-pt}(x',x_2). \notag
\end{align}
In a similar way to (\ref{on1})-(\ref{on4}), we can confirm the cancellation in terms of the integral over the total energy $\epsilon$.  

The on-shell term (\ref{on4}) implies the presence of the following non-covariant term in the effective action 
\begin{align}
\frac{1}{2}\delta C\tilde{\phi}(p^0)^2\tilde{\phi}. 
\label{NC}\end{align}
We have redefined the matter field: $\tilde{\phi}\equiv\Omega\phi$ and focus on the second derivative term. 
The corresponding coefficient $\delta C$ has no time dependence if we fix the physical energy resolution $\Delta E$: 
\begin{align}
\delta C=-\frac{5\kappa^2H^2}{16\pi^2}\int^{\Lambda_\text{R}}_{\Delta E} \frac{d(E-P)}{E-P},\hspace{1em}\Delta E=\Delta\epsilon H|\tau_c|. 
\end{align} 
Note that $P$, $E$ are defined in the same way as in (\ref{PQ}). 
Here $\Lambda_\text{R}$ is the renormalization scale. 

In (\ref{full}), we have assumed that the full propagator respects the covariance. 
Therefore the left-hand side of the Kadanoff-Baym equation (\ref{Dleft}) has no room for the non-covariant contribution (\ref{NC}). 
Of course we do not expect exact Lorentz invariance even for conformally coupled scalar field
since graviton propagators are not scale invariant. Our claim is that the breaking of Lorentz invariance
is small in sub-horizon scale since we can choose $\Lambda_{R}\sim\Delta E$. 

In this argument we have assumed that UV contributions can be renormalized by appropriate counter terms. 
We need to deal with non-covariant divergences of the type (\ref{NC}) due to the non-covariant gauge fixing term (\ref{GF}). 
Namely the matter system with gravity respects the covariance except for the gauge fixing term. 
The non-covariant UV contributions can be absorbed by the matter field redefinitions such as $\tilde{\phi}\to(1+\alpha\tau\partial_0)\tilde{\phi}$, where $\alpha$ denotes an UV divergent constant. 
Since the gauge fixing term is BRS exact, it is in accord with our expectation that
such a non-covariant contribution may be absorbed by a wave function renormalization.

Alternatively we may modify the gauge fixing function in the sub-horizon scale as follows:
\begin{align}
F_\nu^f = D_\mu h^\mu_{\ \nu} - 2 \partial_\nu w -\partial_\mu f(h^\mu_{\ \nu}-2\delta^\mu_{\ \nu}w), 
\end{align}
where $f\rightarrow 2\log a(\tau)$ in the super-horizon scale and $f\rightarrow 0$ in the sub-horizon scale. 
$D_\mu$ is the covariant derivative with respect to the background metric. 
This gauge fixing function smoothly interpolates the original gauge fixing function (\ref{GF}) in the super-horizon scale and the covariant one in the sub-horizon scale.
We exploit the gauge fixing freedom to adopt a manifestly covariant background gauge in the sub-horizon scale.
In this gauge we have verified that the relevant one-loop UV divergence from each diagram is covariant by the standard DeWitt-Schwinger expansion.
Needless to say there is no change with respect to IR logarithms.

We summarize this section. 
When we naively distinguish the off-shell terms from the on-shell terms such as (\ref{on}), (\ref{off}), each term appears to induce the dS symmetry breaking logarithm. 
Such IR singularities originate in the process with a soft or collinear particle. 
Since the off-shell terms are not distinguishable from the on-shell terms in the process, we need to sum up them. 
After the redefinition, the IR cut-off is given by not the initial time but the energy resolution. 
Once the non-local terms are expressed by physical scales such as $\Delta E$, $P$, $\Lambda_\text{R}$, it is apparent that they do not contribute to the dS symmetry breaking. 
We have thus excluded the appearance of the IR logarithms in the non-local terms.
It justifies the hypothesis of the preceding section that the dS symmetry breaking exists only in the local terms.

We also refer to the case that we set the external momentum to be off-shell $p_\mu p^\mu\not=0$. 
In this case, the independence from the initial time can be showed more simply. 
It is because the integral over the total energy is cut-off by the virtuality: 
\begin{align}
\epsilon^2-(p^0)^2>p_\mu p^\mu. 
\end{align} 
The investigation in this section gives a proper interpretation into the on-shell limit of the off-shell effective equation of motion. 

\section{Parametrization dependence}\label{Para}
\setcounter{equation}{0}
 
In this section, we clarify how soft gravitational effects depend on the parametrization of the metric. 
It is the parallel investigation of the previous studies with the effective equation of motion \cite{KitamotoPa}.  
In the previous sections, we investigated soft gravitational effects on a matter system by adopting the parametrization (\ref{para1})-(\ref{para3}). 
As another example, we adopt the parametrization (\ref{Wpara}). 

As explained in (\ref{difference})-(\ref{rename}), the difference between the two parametrizations emerges at the non-linear level. 
So at the one-loop level, the parametrization difference of the metric (\ref{difference}) contributes only to the tadpole diagrams: 
\begin{align}
\Delta(\Sigma_\text{4-pt}(x,x'))=i\delta^{(4)}(x-x')\times\Big[-\kappa\partial'_\mu\big\{\langle h^{\mu\nu}(x')\rangle|_\text{NL}\partial'_\nu\}-\frac{1}{6}\kappa\partial'_\mu\partial'_\nu\langle h^{\mu\nu}(x')\rangle|_\text{NL}\Big], 
\label{tadpole}\end{align}
where $\kappa\langle h_{\mu\nu}(x)\rangle|_\text{NL}$ is identified as
\begin{align}
\kappa\langle h_{\mu\nu}(x)\rangle|_\text{NL}
=-2\kappa^2\langle\Phi(x)\Psi_{\mu\nu}(x)\rangle
-\frac{1}{2}\kappa^2\langle\Psi_{\mu}^{\ \rho}(x)\Psi_{\rho\nu}(x)\rangle
+\frac{1}{8}\kappa^2\langle\Psi_{\rho\sigma}(x)\Psi^{\sigma\rho}(x)\rangle\eta_{\mu\nu}. 
\end{align}
From (\ref{gravityp})-(\ref{P}) and (\ref{rename}), the coefficients of the $\log a(\tau)$ and $a(\tau)$ terms in (\ref{tadpole}) are evaluated as
\begin{align}
\Delta(\Sigma_\text{4-pt}(x,x'))\simeq i\delta^{(4)}(x-x')\times\frac{\kappa^2H^2}{4\pi^2}
\Big\{\log\big(a(\tau')/a_i\big)\big(\frac{3}{4}{\partial'}_0^2+\frac{1}{4}{\partial'}_i^2\big)+Ha(\tau')\big(\frac{3}{4}\partial'_0\big)\Big\}. 
\end{align}
The relative weight of ${\partial'}_0^2$ and ${\partial'}_i^2$ is not equal to $-1$ in the coefficient of the IR logarithm. 
That is, the effective Lorentz symmetry is not respected. 
Since we have confirmed that the effective Lorentz invariance holds in the original parametrization of the metric (\ref{para1})-(\ref{para3}), 
there should be a prescription to retain it in a different parametrization. 

We should recall that the parametrization dependence of the metric emerges only in the tadpole diagrams. 
So we can compensate them by introducing the classical expectation value of the background metric: 
\begin{align}
h_{\mu\nu}\to h_{\mu\nu}+v_{\mu\nu},\hspace{1em}v_{\mu\nu}=-\langle h_{\mu\nu}(x)\rangle|_\text{NL}, 
\end{align}
\begin{align}
\Delta(\Sigma_\text{4-pt}(x,x')) &\to i\delta^{(4)}(x-x')\times\Big[-\kappa\partial'_\mu\big\{\big(\langle h^{\mu\nu}(x')\rangle|_\text{NL}+v^{\mu\nu}\big)\partial'_\nu\} \\
&\hspace{1.2em}-\frac{1}{6}\kappa\partial'_\mu\partial'_\nu\big(\langle h^{\mu\nu}(x')\rangle|_\text{NL}+v^{\mu\nu}\big)\Big]=0. \notag
\end{align}
Note that the gravitational action is stationary with this shift. 
At least at the one-loop level, the compensation by shifting the background metric is available not only for the difference between (\ref{para1})-(\ref{para3}) and (\ref{Wpara}), 
but also for an arbitrary difference of the parametrization of the metric. 
It is because the difference at the non-linear level emerges only in the tadpole diagrams at the one-loop order. 

The discussion in this section is summarized as follows. 
By a judicious choice of the classical expectation value of the background metric, 
the effective Lorentz invariance is preserved for any choice of the parametrization of the metric. 
Furthermore the resulting wave function renormalization factor (\ref{Z}) does not depend on the choice. 

\section{Gauge dependence}\label{Gauge}
\setcounter{equation}{0}

In the preceding sections, we have investigated the IR effects in the gauge condition (\ref{GF}).  
To check whether the obtained results are physical, it is necessary to investigate the gauge dependence of them. 
Although the investigation in this section has some overlaps with our previous paper \cite{KitamotoSD}, we explain it again to make this paper self-sustained. 
Here we introduce a gauge parameter $\beta$ into the gauge fixing term as 
\begin{align}
\mathcal{L}_\text{GF}&=-\frac{1}{2}a^2F_\mu F^\mu, \label{beta}\\
F_\mu&=\beta\partial_\rho h_\mu^{\ \rho}-2\beta\partial_\mu w+\frac{2}{\beta}h_\mu^{\ \rho}\partial_\rho\log a+\frac{4}{\beta}w\partial_\mu\log a. \notag
\end{align}
This gauge fixing term coincides with (\ref{GF}) at $\beta=1$. 
For simplicity, we consider the case $|\beta^2-1|\ll 1$ where the deviation from (\ref{GF}) can be investigated perturbatively. 
The deformation of the action at $\mathcal{O}(\beta^2-1)$ is written as
\begin{align}
\delta \mathcal{L}_{\beta^2-1}=(\beta^2-1)a^4\big[&-\frac{1}{2}a^{-2}(2\partial_\mu w-\partial_\rho h^\rho_{\ \mu})(2\partial^\mu w-\partial_\sigma h^{\sigma\mu}) \label{beta1}\\
&-2H^2(h^{00}-2w)^2+2H^2h^{0i}h^{0i}\big], \notag
\end{align}
where we have neglected the ghost and anti-ghost fields since they do not couple to the matter field. 

Only the minimally coupled modes of gravity contain the scale invariant spectrum and may induce the dS symmetry breaking. 
After neglecting the conformally coupled modes, 
\begin{align}
h^{0i}\simeq 0,\hspace{1em}h^{00}=h^{ii}\simeq 2w, 
\end{align}
the deformation (\ref{beta1}) is reduced to the following form
\begin{align}
\delta \mathcal{L}_{\beta^2-1}\simeq(\beta^2-1)a^2\big[\ 
2\partial_0 h^{00}\partial_0 h^{00}-\frac{2}{9}\partial_i h^{00}\partial_i h^{00}+\frac{2}{3}\partial_i h^{00}\partial_k \tilde{h}^{ki}-\frac{1}{2}\partial_k\tilde{h}^{ki}\partial_l\tilde{h}^{li}\big]. 
\label{beta2}\end{align}
Here the existence of the third and fourth terms indicates that the tensor, vector and scalar modes of the spatial traceless part are not of the same magnitude except for $\beta=1$. 
Up to $\mathcal{O}(\beta^2-1)$ and in the infra-red limit, the nonzero correlation functions of gravity are written as 
\begin{align}
\langle \tilde{h}^T_{ij}(x)\tilde{h}^T_{kl}(x')\rangle&\simeq\int\frac{d^3p}{(2\pi)^3}\ e^{+i{\bf p}\cdot({\bf x}-{\bf x}')}\frac{H^2}{2p^3}\times
(\bar{\delta}_{ik}\bar{\delta}_{jl}+\bar{\delta}_{il}\bar{\delta}_{jk}-\bar{\delta}_{ij}\bar{\delta}_{kl}), \label{beta3}\\
\langle \tilde{h}^V_{ij}(x)\tilde{h}^V_{kl}(x')\rangle&\simeq\int\frac{d^3p}{(2\pi)^3}\ e^{+i{\bf p}\cdot({\bf x}-{\bf x}')}\frac{H^2}{2p^3} \notag\\
&\hspace{2.5em}\times\big(1-\frac{3}{2}(\beta^2-1)\big)(\bar{\delta}_{ik}\frac{p_jp_l}{p^2}+\bar{\delta}_{jl}\frac{p_ip_k}{p^2}+\bar{\delta}_{il}\frac{p_jp_k}{p^2}+\bar{\delta}_{jk}\frac{p_ip_l}{p^2}), \notag\\
\langle \tilde{h}^S_{ij}(x)\tilde{h}^S_{kl}(x')\rangle&\simeq\int\frac{d^3p}{(2\pi)^3}\ e^{+i{\bf p}\cdot({\bf x}-{\bf x}')}\frac{H^2}{2p^3} \notag\\
&\hspace{2.5em}\times\big(1-2(\beta^2-1)\big)3(\frac{p_ip_j}{p^2}-\frac{1}{3}\delta_{ij})(\frac{p_kp_l}{p^2}-\frac{1}{3}\delta_{kl}), \notag\\
\langle h_{00}(x)\tilde{h}^S_{ij}(x')\rangle&\simeq\int\frac{d^3p}{(2\pi)^3}\ e^{+i{\bf p}\cdot({\bf x}-{\bf x}')}\frac{H^2}{2p^3}\times
-\frac{3}{2}(\beta^2-1)(\frac{p_ip_j}{p^2}-\frac{1}{3}\delta_{ij}), \notag\\
\langle h_{00}(x)h_{00}(x')\rangle&\simeq\int\frac{d^3p}{(2\pi)^3}\ e^{+i{\bf p}\cdot({\bf x}-{\bf x}')}\frac{H^2}{2p^3}\times
-\frac{3}{4}\big(1-(\beta^2-1)\big). \notag
\end{align}

Furthermore let us recall that only the local terms contribute to the dS symmetry breaking at least at the one-loop level. 
Namely we may focus on the gravitational propagator at the coincident point. 
As for the propagators at the coincident point, the following replacements are possible
\begin{align}
\frac{p_ip_j}{p^2}\to\frac{1}{3}\delta_{ij},\hspace{1em}\frac{p_ip_jp_kp_l}{p^4}\to\frac{1}{15}(\delta_{ik}\delta_{jl}+\delta_{il}\delta_{jk}+\delta_{ij}\delta_{kl}). 
\end{align}
By considering the replacements, the list (\ref{beta3}) is reduced to the following form
\begin{align}
\langle \tilde{h}^T_{ij}(x)\tilde{h}^T_{kl}(x)\rangle&\simeq\frac{2}{5}(\delta_{ik}\delta_{jl}+\delta_{il}\delta_{jk}-\frac{2}{3}\delta_{ij}\delta_{kl})\frac{H^2}{4\pi^2}\log\big(a(\tau)/a_i\big), \label{beta4}\\
\langle \tilde{h}^V_{ij}(x)\tilde{h}^V_{kl}(x)\rangle&\simeq\big(1-\frac{3}{2}(\beta^2-1)\big)\times\frac{2}{5}(\delta_{ik}\delta_{jl}+\delta_{il}\delta_{jk}-\frac{2}{3}\delta_{ij}\delta_{kl})\frac{H^2}{4\pi^2}\log\big(a(\tau)/a_i\big), \notag\\
\langle \tilde{h}^S_{ij}(x)\tilde{h}^S_{kl}(x)\rangle&\simeq\big(1-2(\beta^2-1)\big)\times\frac{1}{5}(\delta_{ik}\delta_{jl}+\delta_{il}\delta_{jk}-\frac{2}{3}\delta_{ij}\delta_{kl})\frac{H^2}{4\pi^2}\log\big(a(\tau)/a_i\big), \notag\\
\langle h_{00}(x)\tilde{h}^S_{ij}(x)\rangle&\simeq0, \notag\\
\langle h_{00}(x)h_{00}(x)\rangle&\simeq\big(1-(\beta^2-1)\big)\times-\frac{3}{4}\frac{H^2}{4\pi^2}\log \big(a(\tau)/a_i\big). \notag
\end{align}
It can be simply rewritten as 
\begin{align}
\langle h_{\mu\nu}(x)h_{\rho\sigma}(x)\rangle\simeq\big(1-(\beta^2-1)\big)\times P_{\mu\nu\rho\sigma}\frac{H^2}{4\pi^2}\log \big(a(\tau)/a_i\big),  
\label{beta5}\end{align}
where $P_{\mu\nu\rho\sigma}$ is defined in the same way in (\ref{P}). 
As a consequence, the gravitational propagator in the deformed gauge condition (\ref{beta}) is proportional to the propagator in the original one (\ref{GF}) at the coincident point. 

From (\ref{beta5}), it turns out that at least up to $\mathcal{O}(\beta^2-1)$, the effective Lorentz invariance is respected for a continuous gauge parameter $\beta$.  
Namely the soft graviton effect emerges as the following overall factor
\begin{align}
Z_\beta(\tau_c)=1-\big(1-(\beta^2-1)\big)\frac{3}{8}\frac{\kappa^2H^2}{4\pi^2}\log\big(a(\tau)/a_i\big). 
\label{beta6}\end{align}
Although it depends on a gauge parameter, such an overall factor can be absorbed by the wave function renormalization in a similar way to (\ref{WR}):  
\begin{align}
\tilde{\phi}(x)\to Z^\frac{1}{2}_\beta(\tau)\tilde{\phi}(x),\hspace{1em}\langle\tilde{\phi}(x_1)\tilde{\phi}(x_2)\rangle\to Z^{-1}_\beta(\tau_c)\langle\tilde{\phi}(x_1)\tilde{\phi}(x_2)\rangle. 
\label{beta7}\end{align}
That is, there is no physical effects from soft gravitons in the free scalar field theory (\ref{action}). 
We have confirmed that the effective Lorentz invariance is respected also in the Dirac and gauge field theories \cite{KitamotoSD,KitamotoG}. 

Furthermore we refer to the soft gravitational effects in the interacting matter system. 
If the matter action contains interaction terms, they are modified after the wave function renormalization. 
In addition to the wave function renormalization contributions, the interaction terms are corrected by soft gravitons dressing the vertices. 
We have found that the couplings of the quartic, Yukawa and gauge interactions are screened by soft gravitons \cite{KitamotoSD,KitamotoG}.  

As seen in (\ref{beta7}), the wave function renormalization factors depend on a gauge parameter and the same is true for the vertex contributions. 
That is, the time dependence of each coupling is gauge dependent as
\begin{align}
\delta g_j\sim \beta_j t, 
\end{align} 
where $g_j$ denotes each coupling and the corresponding coefficient $\beta_j$ depends on a gauge parameter. 
It may be expected as time is an observer dependent quantity. 
Our proposal is to pick a particular coupling as a physical time such as $g_i\sim t$. 
We then measure the variation speed of the other couplings in terms of it
\begin{align}
\delta g_j\sim \frac{\beta_j}{\beta_i}t. 
\end{align}
We have argued that this idea works as the relative ratio of the variation speed of the couplings is invariant under an infinitesimal change of gauge, 
specifically up to $\mathcal{O}(\beta^2-1)$. 

A nice analogy we can draw here is 2-dimensional quantum gravity. 
Of course there is no gravitons in 2-dimensional gravity. 
Nevertheless we can measure the scaling dimensions of the local operators. 
It is equivalent to measure the scaling dimensions of the couplings to these local operators. 
These couplings are found to acquire nontrivial scaling dimensions due to quantum fluctuation of the metric 
\begin{align}
\delta f_j\sim \mu^{\alpha_j}, 
\end{align}
where $\mu$ denotes a scale in a particular gauge. 
The scaling dimension of each coupling is gauge dependent. 
It makes sense that there is no unique way to specify the unit of scaling in quantum gravity. 
This ambiguity is resolved by picking a particular coupling to specify the unit of scaling dimension such as $f_i\sim \mu$. 
In this normalization, the other couplings scale as
\begin{align}
\delta f_j\sim \mu^\frac{\alpha_j}{\alpha_i}. 
\end{align}
Indeed the relative ratio of scaling dimensions are gauge independent in 2-dimensional quantum gravity \cite{KN,KKN}. 
In the conformal gauge, the scale factor of the metric (conformal mode) is the only degrees of freedom except for the ghost and anti-ghost fields. 
The dynamics of the conformal mode is described by the Liouville theory and it explains the scaling dimensions of the operators. 
The contribution from the the conformal mode is not canceled by that from the ghost and anti-ghost fields. 
That is why the scaling dimension of each coupling is gauge dependent. 

In 4-dimensional dS space, we have parametrized the metric by the conformal mode and traceless modes $h^\mu_{\ \nu},\ h^\mu_{\ \mu}=0$. 
The traceless modes consist of $h^{00}=h^{ii}$, $h^{0i}$ and the spatial traceless modes $\tilde{h}^i_{\ j},\ \tilde{h}^i_{\ i}=0$. 
The spatial traceless modes can be further decomposed into the scalar, vector and tensor modes. 
Some of them: $h^{0i}$ and $X$ consisting from $h^{00}$ and the conformal mode are with negative norms. We can decompose the whole propagator into the contributions from respective modes. 
Although the propagator due to graviton (spatial traceless and transverse mode) is gauge independent, the rest of the propagator is gauge dependent. 
The time dependence of the couplings arises due to the whole metric degrees of freedom and the contribution from the ghost and anti-ghost fields can be neglected in our gauge. 
That is why they are gauge dependent. Nevertheless we argue that the gauge dependence should be canceled in physical observables such as Lorentz invariance and the ratio of the variation speeds of the couplings.

\section{Conclusion}\label{Conclusion}
\setcounter{equation}{0}

Due to the existence of the scale invariant spectrum, 
we need to introduce an IR cut-off into the propagator for a massless and minimally coupled scalar and gravitational field. 
The IR cut-off fixes the minimum value of the comoving momentum and it is identified as an inverse of the initial time. 
As a consequence, the propagator has a logarithmic dependence of the scale factor which breaks the dS symmetry. 

It should be noted that there is another IR contribution in massless field theories. 
In the process with a soft or collinear particle,  the frequency of the integrand becomes small and  so the integral over the negatively large conformal time is dominant. 
When we set the external momentum to be off-shell, the time integral is bounded by not the initial time but an inverse of the virtuality. 

In the on-shell limit, IR singularities occur since the frequency can vanish. 
However we cannot distinguish the off-shell term from the on-shell term in the zero frequency process. 
The IR singularities cancel after summing over degenerate states between real and virtual processes. 
The corresponding IR cut-off is given by the energy resolution. 
We observe phenomena in the condition where the physical scale of the virtuality or the energy resolution is fixed. 
Therefore the non-local contribution respects the dS symmetry. 

The above cancellation originates from the fact that the total spectrum weight is preserved. 
In this regard, it may be a universal phenomenon as far as field theoretic models are consistent with unitarity. 
By using the Kadanoff-Baym approach, we have specifically shown that the cancellation holds in a matter system with gravity. 
That is, soft gravitons contribute to the dS symmetry breaking only through the local terms. 
In other words, if the dS symmetry breaking takes place, it originates only in the increasing degrees of freedom at the super-horizon sale. 
It has justified the assumption adopted in \cite{KitamotoSD}.  

We have found that the local contribution appears as a time dependent overall factor of the kinetic term. 
It indicates that the effective Lorentz symmetry is respected even if soft gravitational effects are considered. 
After the wave function renormalization, soft gravitons leave no growing physical effect to a free matter system. 
Of course, the wave function renormalization factor is the same one obtained in \cite{KitamotoSD}. 

Furthermore we have investigated how soft gravitational effects depend on the parametrization of the metric. 
The parametrization dependence appears only in the tadpole diagram. 
Thus it can be compensated by introducing the classical expectation value of the metric. 
With this prescription, the gravitational action is kept stationary and the results obtained in this paper do not depend on the parametrization of the metric. 
It is consistent with the previous studies based on the effective equation of motion \cite{KitamotoPa}. 

When we consider a slightly deformed gauge condition, the soft graviton effect on the free field theory depend on a gauge parameter.  
However the gauge dependent IR effect respects the effective Lorentz invariance and emerges just as an overall factor. 
Thus soft gravitons do not contribute to the free field theory even in the deformed gauge condition. 

On the other hand, soft gravitons give rise to important physical effects in interacting field theories. 
The effective couplings become time dependent although each variation speed is gauge dependent. 
We argue that the gauge dependence is due to the fact that time is an observer dependent quantity.  
As an analogy of 2-dimensional quantum gravity, we can measure the variation speeds of the couplings gauge invariantly by their relative scaling exponents.  
Of course, it is still an open problem whether the effective Lorentz invariance and the gauge invariance of the relative scaling exponents hold or not against the large gauge deformation.   

The investigation in this paper is on the two-point function. 
In the previous studies \cite{KitamotoSD}, we investigated soft gravitational effects in interacting field theories. 
There we assumed that only the local terms contribute to the dS symmetry breaking also in multi-point functions. 
It remains an open problem whether the statement is correct or not. 
However we believe that the conjecture is reasonable since the cancellation of the non-local IR singularities is intimately connected to the unitarity of the theory. 

\section*{Acknowledgment}
This work is supported in part by the Grant-in-Aid for Scientific Research 
from the Ministry of Education, Science and Culture of Japan, 
and the National Research Foundation grants NRF-2005-0093843, NRF-2010-220-C0000, NRF-2012KA1A9055. 


\end{document}